\documentclass{iaus}
\usepackage{graphics}

  \checkfont{eurm10}
  \iffontfound
    \IfFileExists{upmath.sty}
      {\typeout{^^JFound AMS Euler Roman fonts on the system,
                   using the 'upmath' package.^^J}%
       \usepackage{upmath}}
      {\typeout{^^JFound AMS Euler Roman fonts on the system, but you
                   dont seem to have the}%
       \typeout{'upmath' package installed. iaus.cls can take advantage
                 of these fonts,^^Jif you use 'upmath' package.^^J}%
      }
  \else
  \fi

  \checkfont{msam10}
  \iffontfound
    \IfFileExists{amssymb.sty}
      {\typeout{^^JFound AMS Symbol fonts on the system, using the
                'amssymb' package.^^J}%
       \usepackage{amssymb}%

      }{}
  \fi

  \IfFileExists{amsbsy.sty}
    {\typeout{^^JFound the 'amsbsy' package on the system, using it.^^J}%
     \usepackage{amsbsy}}
    {}

\newsavebox{\astrutbox}
\sbox{\astrutbox}{\rule[-5pt]{0pt}{20pt}}

\title[The Interplay among Black Holes, Stars and ISM in Galactic 
       Nuclei]{The role of bars}

\author[F. Combes]%
{F. Combes}

\affiliation{Observatoire de Paris, LERMA,
61 Av. de l'Observatoire, F-75014, Paris, France 
 email:francoise.combes@obspm.fr}

\pubyear{2004}
\volume{222}
\pagerange{1--8}
\date{?? and in revised form ??}
\jname{The Interplay among Black Holes, Stars and ISM \\in Galactic Nuclei}
\editors{Th. Storchi Bergmann, L.C. Ho \& H.R. Schmitt, eds.}
\begin{document}

\maketitle

\begin{abstract}
Secular evolution and fueling is driven by bars in spiral galaxies, 
and the related dynamical processes
are reviewed. It is shown that gravity torques dominate over viscous torques, and 
produce gas infall to the center. In this infall, the bar wave accepts the angular
momentum, which produces its destruction. In the end, a central mass concentration
is build, which also contributes to this destruction. While gas can be stalled at ILR
for a while, secondary bars then take over the fueling. In a galaxy life-time, several bar
episodes can successively develop. Through external gas accretion, 
disks are replenished by intermittence, between two bar phases. While the general trend for a
galaxy is to evolve towards early-types, its morphology can oscillate, and turn back
transiently to later-types.
\end{abstract}

\firstsection 

\section{The problem of bars and AGN fueling}

Disks of spiral galaxies can be considered as accretion disks: matter
tends to concentrate to reach the least energy state, but the angular momentum
has to be transfered outwards. Since viscous torques are not sufficient over
a Hubble time, gravity torques play the main role, through formation of bars and
spirals. Numerical simulations reveal that bars are very efficient to 
drive the gas of a galaxy inwards, however there is no clear correlation
between nuclear activity and bars in the observations
(e.g. Mulchaey \& Regan 1997). There could even be a
lower fraction of strong bars in Seyferts (Shlosman et al 2000).
The only positive signs of the bar action is the observation of 
more outer rings in Seyfert galaxies (Hunt \& Malkan 1999, and Ag\"uero et al, this meeting),
and more gas concentration in barred galaxies (Sakamoto et al 1999).
Also, there are always circumnuclear dust in AGN, while not 
always in normal galaxies (Martini et al 2003).

The main reason for this lack of correlation could be the different
time-scales: the duty cycle for nuclear activity is of the order of
40 Myr,  much shorter than the dynamical time-scales of gas 
inflow (at least 300 Myr). The case of galaxy interactions is similar,
since internal bars are then also the motor of gas inflow.
Nuclear starbursts are often simultaneously observed
with nuclear activity (cf Maia, Lim et al, Storchi-Bergman et al,
this meeting).

\section{Gas flows in barred galaxies}

Gas behaviour in a barred potential has been studied for a
long time (e.g. Sanders \& Huntley 1976), but the larger spatial
resolution reached recently have allowed to better estimate
the influence of viscosity (Maciejewski et al 2002). Already 
Englmaier \& Gerhard (1997) revealed the large influence of
the sound speed in the gas behaviour. The larger the
velocity dispersion in the gas, the larger the mean free path
of the particles, and the larger the distance over which they can exchange
angular momentum, so that shear is more important. With large
velocity dispersion, the gas that was previously trapped at ILR
can further fall inwards (Maciejewski et al 2002,
cf also Ann, this meeting).
The role of bar characteristics (strength, axis ratio, etc..)
on gas inflow has been examined in details
(Regan \& Teuben 2004), weak bars have only marginal effect,
while gas velocity dispersion or sound speed Vs is a key parameter.
Bars drive the gas until the ILR only, if Vs is small.
If Vs is larger than 20km/s, the viscous shear can drive the gas inwards.

\subsection{Influence of gas physics}

The multi-phase interstellar medium is frequently over-simplified, 
and simulated in 2D. Wada \& Norman (2001) have shown part of the
hidden conplexity in  detailed 3D gas simulations in a fixed 
spherical potential (with dark matter, stars, central black hole).
SN explosions are followed to heat the gas, which scale height
increases with the square root of star formation rate. Even with large
viscosity and velocity dispersion, the gas flow rate towards the center
is low ($<$ 1M$_\odot$/yr). However, there is no bar in these models.

The fact that viscous torques do not predominate in galaxy disks
is supported by the observations of resonant rings, in particular
the outer rings, due to the presence of positive gravity torques 
between corotation and OLR (Schwarz 1981). Also the large
frequency of nuclear rings, where gas accumulates and form stars
(Buta \& Combes 1996, 2000) gives the idea of a low sound speed of gas,
with low velocity dispersion.

\subsection{Gravity versus viscous torques}

In numerical simulations, several measurements can be
done to test the importance of viscous relative to gravity 
torques. When there is no bar, the gas is not inflowing,
which means that viscous torques alone are insufficient.
But they could play a role when there is a bar, since
orbits are no longer circular, and the shear could then be
higher. It is the gas dissipation that provokes the phase shift
between gas and stars, and allows gravity torques to play a larger
role. In models where gas is represented by sticky particles, 
it is straightforward to compute the angular momentum exchanged
in cloud collisions, and compare to the gravity torques.
Such a comparison (Bournaud \& Combes 2004) is displayed
in Figure 1, where it is clear that the inflowing gas  
loses its angular momentum only by gravity torques.

Measurements of gravity torques from the observations confirm
that the importance of gas inflow is not over-estimated in
the simulations; the phase shift of the characteristic dust-lanes,
offset towards the leading edge of the bar, is of the same magnitude, and 
the torques computed from the red images are of the same order
as the computed ones (e.g. Block et al 2002).  

\begin{figure}
\centerline{\includegraphics{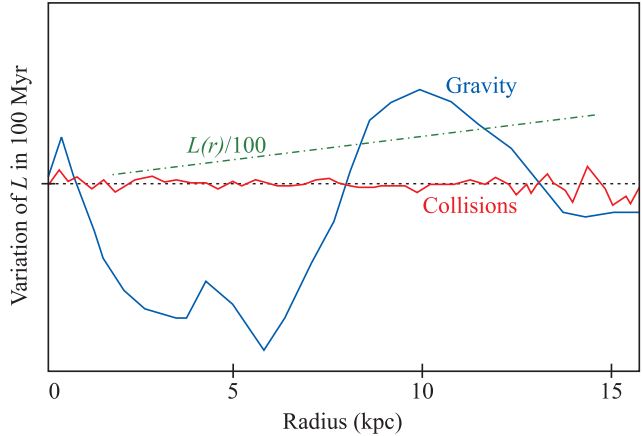}}
\caption{Comparison of the angular momentum lost in cloud collisions
(viscous torques), and that due to gravity torques
(from Bournaud \& Combes 2004). Gravity torques change sign at corotation.}
\end{figure}

\section{Destruction of bars}

The consequence of the gas inflow towards the center is the
formation of a central mass concentration (CMC) 
and the destruction of the bar. This was shown 
both by N-body simulations, and orbit computations in fixed potentials
(Hasan et al 1990, 1993, Combes 1994, Norman et al 1996, 
Berentzen et al 1998).
Bars are destroyed by  1-5\% mass concentrations within 1kpc
(percentage with respect to the disk mass).
According to the hardness of the CMC,
orbits sustaining the bar (x1) are scattered by a black hole,  
and become chaotic, or bars are
weakened or dissolved, by a concentrated bulge or nuclear disk.
Destruction is more efficient for a more concentrated CMC
(a true black hole of 0.5\% mass is sufficient, Hozumi \& Hernquist 1999)
and also if the growth-time of the CMC is short.

Recently Shen \& Sellwood (2004) contest the fragility of bars,
since usual black hole masses in spiral galaxies are not large enough,
and softer gas concentrations only weaken the bars. They use however a non
self-consistent galaxy model, with an artificial CMC growth in the center,
without gas.  They do not consider how the CMC is formed, which is in itself
the main destruction mechanism.

\subsection{Role of gas in bar destruction}

One essential factor in the bar destruction, that was
overlooked until now, is the angular momentum loss of the
infalling gas to the bar wave (Bournaud \& Combes 2004).
Gas is driven in by the bar gravity torques, and
its angular momentum is taken up by the bar wave.
Since the bar, inside its own corotation,
 has negative angular momentum, this destroys the bar
(the loss of angular momentum from the 
gas infalling from corotation to center is of the same order).

Figure 2 compares several N-body experiments with gas
and star formation, to probe the role of various 
phenomena. The gas loss of
angular momentum is sufficient to destroy the bar completely,
and this occurs even before a CMC is built. The 
presence of a CMC of only 1\% is not enough to destroy the bar,
but 1-2\% of gas infall is enough, which explains the result
of previous simulations (Berentzen et al 1998, 
Bournaud \& Combes 2002, 2004).

Since the main reason of bar destruction is not the presence
of a CMC, it will then be more easy to re-form a bar later.

\begin{figure}
\centerline{\includegraphics{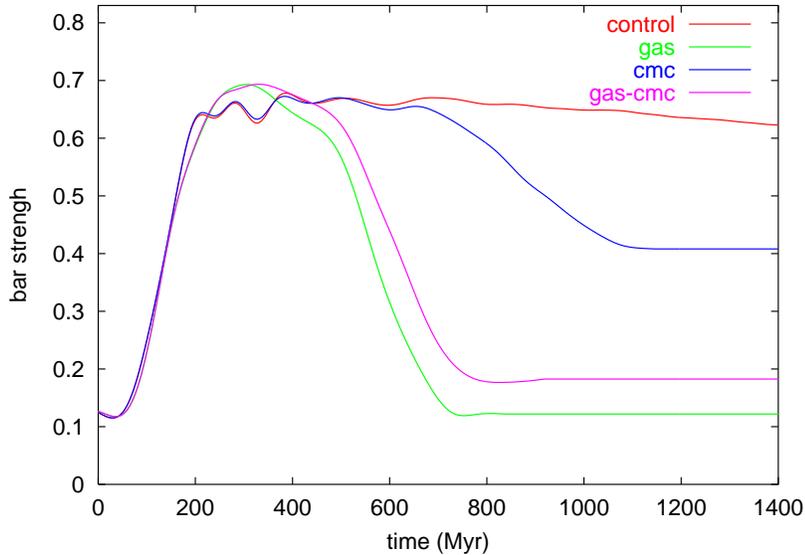}}
\caption{Strength of the bar as a function of time, for the control run
at the top (stellar disk without gas), for the realistic model (light curve
at bottom), and the same but where the CMC is progressively removed (full
curve, ``gas-cmc'', almost coincident). When only an equivalent CMC is introduced, the bar
is only weakened (intermediate curve). The model galaxy had 6\% of mass in gas with respect 
to the disk, a bulge of 25\%, and 1\% of disk mass in gas has moved inside 300pc at the
end.}
\end{figure}

\subsection{Double bars and gas inflow}

Nested bars are found in N-body simulations to decouple from the primary bar when 2
ILR exist, and consequently perpendicular x2 orbits. The nuclear and primary bars
have one resonance in common, which avoids the chaos, and allows
the coupling and exchange of energy (Friedli \& Martinet 93).

The presence of this second bar, which torques prolonge that of the primary bar,
is essential for gas inflow. Without it, the gas may be stalled at ILR.
The only way to drive it to the center is through viscous torques,
in the case of strong shocks and large velocity dispersion
(Maciejewski 2002). Another possibility, if the gas accumulates and is cold, is
clump instability; giant clouds will lose angular momentum through
dynamical friction against the bulge.

\section{Bar reformation}

Since bars in spiral galaxies with gas are easy to destroy, they must be
regularly reformed, to explain the observed bar frequency today.
The typical destruction time-scale is 1 Gyr, as is the formation time-scale.

\subsection{Statistics on bar strength}
  It is commonly adopted from galaxy classification that about
two thirds of galaxies are barred (SB and SAB), while one third
is strongly barred (SB). This percentage corresponds to visible
bands, there are even more barred galaxies in near-infrared images.
To quantify better bar strength in spiral galaxies, and compare to
N-body models, we have undertaken to compute the parameter $Q_b$
for the near-infrared OSU survey of 163 galaxies (Eskridge et al 2002).

The bar strength parameter $Q_b$ is estimated from
the Fourier decomposition of the potential, it is the ratio of the $m=2$ component
of the tangential force to the radial force. The potential is estimated
from the NIR images after disk deprojection, and assuming a constant M/L
  (Block et al 2002). The main result is a dearth of axi-symmetric galaxies,
and a large number of strongly barred galaxies
(see also Whyte et al 2002, whose estimation relies on bar axis ratio).

This observed distribution of bar strength cannot be accounted for
in numerical models of isolated galaxies. After a few Gyr, galaxies
become axisymmetric, or weakly barred. The only possibility to reform
the bar is gas accretion from outside, to replenish the disk, decrease
the bulge-to-disk mass ratio, and make the disk unstable again.
Numerical simulations can quantifify more precisely the accretion rate,
by confrontation to the observed bar strength distribution.

\subsection{Numerical simulations}

Since accretion of large quantities of gas have to be considered,
star formation is essential to follow the fate of this gas.
Simulations will show that the gas fraction remains almost
constant during evolution, the rate of gas accretion being 
comparable to the rate of star formation (roughly 10 M$_\odot$/yr).
 
Also, the required gas could come from the mass loss from stars,
and it is important to take into account the mass loss quantitatively,
via a continuous recycling, and not instantaneous, scheme 
(e.g. Jungwiert, Combes \& Palous 2001).

Gas accretion simulations have been carried out with a 3D FFT 
N-body code, and a sticky particles model for gas clouds
(Bournaud \& Combes 2002).
Star formation is followed according to a Schmidt law recipe, 
with exponent 1.4.
The morphology of simulated spiral galaxies change considerably
in the long term
when gas accretion is taken into account. An isolated galaxy first
forms a bar, and the gas is driven inwards. The morphology corresponds
to a normal spiral, but only for ~ 1 Gyr. Then the bar is destroyed, and 
the galaxy turns axisymmetric. With accretion, the galaxy is re-juvenated:
the disk becomes unstable again to bars and spirals. Long term
evolution requires gas accretion, in order to obtain the presently observed
spiral morphologies.

Since the bar exerts positive torques on the gas from corotation
until OLR, the accreted gas has to wait for the destruction of the bar
to replenish the disk. During a galaxy life-time, gas accretes by intermittence.
First it is confined outside OLR, until the bar weakens, through gas infall
from corotation to the center. Then the galaxy becomes axisymmetric, and
only through viscosity the external gas settles in the disk,
to make it unstable again to bar formation

Given the time-scales to bar destruction and formation, it is
possible to expect 3 or 4 bar episodes in a galaxy life-time.
Simulations show that
the successive bar pattern speeds increase with time (the corotation
radius shrinks).
It is interesting to note that, in this evolution, the
bulge-to-disk mass ratio can decrease, and therefore the galaxy
can evolve towards later types, for a short period. While 
the usual trend is to evolve towards early-types, 
the morphological type of a spiral galaxy can oscillate.

\subsection{Origin of the gas}

Numerical simulations with accretion predict the bar frequency
today, as a function of the accretion rate. When compared to
the observed frequency of bars (see above), a large accretion rate
is required, such that a galaxy doubles its mass in 10 Gyr. For
a galaxy of the Milky Way type, this corresponds to an
accretion rate of the order of 10 M$_\odot$/yr.

Could this gas be accreted through galaxy interactions?
Massive galaxy interactions and mergers do not develop
disks but spheroids. Gas-rich dwarf companions cannot
provide more than 10\% of the accretion, since those
interactions heat the galaxy disk (e.g. Toth \& Ostriker 1992).
What is required is a continuous source of cold gas, which could
come from the cosmic filaments in the near environment of galaxies.

Cosmological accretion is comparable to what is
required for bar reformation. This
accretion is compatible with doubling the mass in 10 Gyr.
It is able also to explain the history of star 
formation in typical spiral galaxies (Semelin \& Combes 2002).
 The accretion is intermittent, rythmed by gravitational
instabilities; the gas infalls also fuels the central
black hole, which grows in parallel to the bulge
(cf Combes 2000).

\section{Conclusions: fueling processes}

When gravity torques exist, they are the most efficient
mechanism to fuel gas towards the center. The unstable
disk forms first a primary bar, then a secondary bar 
decouples, but is more short-lived: ILR can stop inflow 
only for a while.
When the gravity torques disappear, then:
-- mild gravitational instabilities develp velocity
dispersion and  viscosity

-- or create clumps of gas; the disk then very non-axisymmetric 
produces inflows

-- or dynamical friction of giant clouds against the bulge
brings the gas to the center,

-- alternatively $m=1$ asymmetries, due to a companion, or anisotropic accretion,
accelerate infall (examples in NUGA survey, cf Garcia-Burillo, this meeting).

Numerical simulations have clearly shown how all 
these dynamical processes are self-regulated. It is
 unrealistic to study phenomena independently, like 
gas flow in rigid bars for
instance, since the gas infall regulates the 
bar strength. {\bf Fueling bars self-destroy.}

Galaxies are not closed systems, they accrete gas
by intermittence, and double their mass every 10 Gyr.
This accretion cannot come mainly from companions,
but come from gas in the near cosmic filaments.
This high accretion rate 
is required to explain bar/spiral reformation, and
also star formation.

\end{document}